\def \m{\mu}

\def \hg {{\hat g}}

\newcommand{\rf}[1]{(\ref{#1})}
\newcommand{\beq}{\begin{equation}}
\newcommand{\eeq}{\end{equation}}
\newcommand{\bea}{\begin{eqnarray}}
\newcommand{\eea}{\end{eqnarray}}
\newcommand{\beas}{\begin{eqnarray*}}
\newcommand{\eeas}{\end{eqnarray*}}

\documentstyle[12pt,epsf]{article}
\baselineskip=\normalbaselineskip
\def \m{\mu}

\ifx\TwoupWrites\UnDeFiNeD\else\target{\magstepminus1}{11.3in}{8.27in}
	\source{\magstep0}{7.5in}{11.69in}\fi
\newfont{\fourteencp}{cmcsc10 scaled\magstep2}
\newfont{\titlefont}{cmbx10 scaled\magstep3}
\newfont{\authorfont}{cmcsc10 scaled\magstep1}
\newfont{\fourteenmib}{cmmib10 scaled\magstep2}
	\skewchar\fourteenmib='177
\newfont{\elevenmib}{cmmib10 scaled\magstephalf}
	\skewchar\elevenmib='177
\makeatletter
\newcommand\nonsequentialeqnum{
	\@addtoreset{equation}{section}
	\def\theequation{\arabic{section}.\arabic{equation}}}
\newif\ifp@bblock  \p@bblocktrue
\newcommand\nopubblock{\p@bblockfalse}
\newcommand\topspace{\hrule height 0pt depth 0pt \vskip}
\newcommand\p@bblock{\begingroup \tabskip=\hsize minus \hsize
	\baselineskip=1.5\ht\strutbox \topspace-2\baselineskip
	\halign to\hsize{\strut ##\hfil\tabskip=0pt\crcr
	\the\Pubnum\crcr\the\date\crcr}\endgroup}
\renewcommand\titlepage{\ifx\TwoupWrites\UnDeFiNeD\null\vspace{-1.7cm}\fi
\vskip0.6cm
	\ifp@bblock\p@bblock \else\hrule height 0pt \relax \fi}
\makeatother
\newtoks\date
\newtoks\Pubnum
\newtoks\pubnum
\Pubnum={
FIT-HE-96-81 \crcr
hep-th/xxxxxxxx
}
\date={\today}
\newcommand{\frontpageskip}{\vspace{12pt plus .5fil minus 2pt}}
\renewcommand{\title}[1]{\frontpageskip
	\begin{center}{\titlefont #1}\end{center}\par}
\renewcommand{\author}[1]{\frontpageskip\par\begin{center}
	{\authorfont #1}\end{center}
	\nobreak
	}

\newcommand{\address}[1]{\par\begin{center}{\sl #1}\end{center}\par}

\renewcommand{\thanks}[1]{\footnote{#1}}
\renewcommand{\abstract}{\par\frontpageskip\centerline{\fourteencp Abstract}
	\vspace{8pt plus 3pt minus 3pt}}
\begin{document}
\pubnum{93-37}
\date{August 1996 \crcr}
\titlepage

\renewcommand{\thefootnote}{\fnsymbol{footnote}}
\title{
String Loop Corrections and the Condensation of\\
   Tachyon in 2d Gravity\\
}

\author{
K.\ Ghoroku \thanks{
e-mail address: gouroku@dontaku.fit.ac.jp}
}

\address{

Fukuoka Institute of Technology\\
Wajiro, Higashi-ku, Fukuoka 811-02, Japan \\
}

\renewcommand{\thefootnote}{\arabic{footnote}}
\setcounter{footnote}{0}
\newcommand{\cleqn}{\setcounter{equation}{0} \indent}
\renewcommand{\theequation}{\thesection.\arabic{equation}}
\newcommand{\beqa}{\begin{eqnarray}}
\newcommand{\eeqa}{\end{eqnarray}}
\newcommand{\eq}[1]{(\ref{#1})}

\begin{abstract}
  Quantum theory of 2d gravity is examined by including a special
quantum correction, which corresponds to the open string loop corrections
and provides a new conformal anomaly for the corresponding $\sigma$
model. This anomaly leads to the condensation of the tachyon, and 
the resulatnt effective theory implies a possibility of extending
the 2d gravity to the case of $c>1$.

~\\
~\\
\end{abstract}

\newpage

\renewcommand{\thesection}{\arabic{section}.} 
\renewcommand{\theequation}{\thesection \arabic{equation}}
\newcommand{\scs}{\setcounter{equation}{0} \setcounter{section}}
\def\req#1{(\ref{#1})}
\setcounter{footnote}{0}

\section*{1. Introduction}\scs{1}

Two dimensional gravity
is closely related to the non-critical string theory in 
the sence that the string
world sheet action can be identified with the 
quantized 2d gravitational theory. In both theories, the vacuum state
is determined by the principle of the confromal invariance.
In spite of the large development of 2d gravitational theory,
there remains a problem how to add a number of matter fields in a way
not to destroy the stability of the surface. 
The study of the vacuum state of the 2d surface so far has been performed
through the field equations obtained from the short distance behavior
of the fields living on the surface, and
we can find the complex dressed factor of the tachyon-perturbation for $c>1$
\cite{ddk}, where $c$ denotes the central charge.
This complex dressed factor
implies the instability of the vacuum 2d gravity since the tachyon is
corresponding to the cosmological term in the 2d action.
This difficulty is known
as the $c=1$ wall problem, and its origin is reduced to 
the appearance of the true tachyon mode, which leads to the unstable
surface fluctuation.
Although 
the properties like the renormalization group equations of matter fields
are not sensitive to the vacuum state of the surface \cite{poly},
we should reconsider the string field equations
obtained by a simple
analysis of $\alpha'$ expansion \cite{clty} in order to resolve the
problem of the vacuum of 2d surface.

In order to approach this problem, we consider two
effects other than the usual $\alpha'$-expansions. One is the
nonlinear term of the tachyon equation. This is obtained by expanding
the tachyon field in its power series and picking up the anomaly
~\cite{ito,coop}.
This method is nonperturbative from the viewpoint of $\alpha'$-expansion, 
but it would be difficult to resolve $c=1$ problem
by this term only ~\cite{tsey,bank}.
Another is the effect of the fluctuations of the surface
which could be formulated by the string loops \cite{fis}.
The most simple one is obtained
by the one-loop correction of open string whose end
point is connected to the boundary on the surface
\cite{clny}. This state of the
surface can be constructed as a boundary state with a tube of the
closed string which propagates starting from the surface into the
vacuum. This formulation has been given for the case
of the superstring or for the critical string, and its 
technique ~\cite{clny,fta}
has recently attracted many attentions since it
is applied to the derivation of the D-brane action \cite{schm}.
Our purpose is here to
extend this formulation to the case of
the noncritical string to resolve the problem of 2d gravity mentioned
above.

The analysis is performed near the 
linear dilaton vacuum, and we assume that the tachyon field 
is small and the string
coupling constant is also small 
for the sake of the consistency of the approximation.
The guiding principle of our analysis
is the conformal invariance or the BRST invariance of the theory.
In the estimation of the anomaly for the boundary state with a tube
in the noncritiacl string theory,
we should notice the following two points. (i)
This tube amplitude badly diverges for $c>1$,
where the tachyonic state appears under the linear dilaton vacuum. 
(ii) The second point is the appearance of the
Liouville field and the
background charge term in the world sheet action.

In order to evade the first point, we consider for
$c\leq 1$. 
And it might be possible to approach the theory of $c>1$ from the analysis
of $c=1$ case. In fact, we might provide a clue to the extension to 
this direction. To resolve the second point in our analysis,
we propose a small device of the technique so that it is applicable to
the non-critical string theory and to the 2d gravity.

\section*{2. Tree Vacuum State}\scs{2}

We set up a vacuum state to calculate the string loop
corrections. Such a
vacuum for the string 
theory is obtained from the principle of the conformal invariance of 
its world-sheet action. Equivalently,
the effective action of the quantized 2d gravity is also given by 
imposing the same principle on
the following non-linear $\sigma$-model form of the action,
\bea
             S_{\rm 2d}&=&{1 \over 4\pi}\int\,d^2z\sqrt{\hg}
            \left[ {1 \over 2}G_{\mu\nu}(X)\hg^{\alpha\beta}
              \partial_\alpha X^{\mu}\partial_\beta X^{\nu}
           +{\hat R}\Phi(X)+T(X) \right]+\hat{S}_{\rm gh} , 
                           \label{eq:t1} \\
           \hat{S}_{\rm gh} &=& {1 \over 2\pi}\int\,d^2z\sqrt{\hg}
              \hat{g}^{\alpha\beta}c^{\gamma}\nabla_{\alpha}
              b_{\beta\gamma}\ \ , \label{eq:t11}
\eea
where the conformal gauge is taken, 
$g_{\alpha\beta}=\rm{e}^{2\phi}\hat{g}_{\alpha\beta}$ and 
$\hat{g}_{\alpha\beta}$ is some
fiducial metric. The conformal mode ($\phi$) and $c$-scalar fields
($x^i$) in the world sheet action are denoted
by $X^{\mu}=\left\{\phi, x^i\right\}$, where $\mu=0,i$ and
$i=1 \sim c$. $\hat{S}_{\rm gh}$ represents the ghost-action.
The result of $\alpha'$-expansion and $T$-expansion so far 
is summarized by the following target space action,
\beq
 S_{\rm T}={1 \over 4\pi}\int d^dX\sqrt{G}{\rm e}^{-2\Phi}
          \left\{R-4(\nabla \Phi)^2+(\nabla T)^2+v(T)-\tilde{k}
            \right\}, \label{eq:t2}
\eeq
where $\tilde{k}=(25-c)/3$, $d=1+c$ and
\beq
 v(T)=-2T^2+{1 \over 6}T^3+\cdots. \label{eq:t3}
\eeq
The second term of \rf{eq:t3}
has been obtained by expanding the tachyon term,
$\int\,d^2z\sqrt{\hg}T(X)$, of \rf{eq:t1} in its series
~\cite{ito,coop}. In each Feynman graph of this expansion,
infinite series of $\alpha'$
are included since the fluctuation field is exponentiated in the vertex.
Then this expansion is regarded as a
non-perturbative procedure from the viewpoint of the $\alpha'$-expansion.
Here, we consider this term 
simultaneously with the perturbative terms obtained by the $\alpha'$-expansion
since it is non-derivative one. 
Other higher order terms of $T$ (the dotted part) are suppressed
here by assuming $T<<1$.

From \rf{eq:t2}, the following
linear dilaton vacuum is obtained as a
well defined vacuum of the theory at tree level,
\beq
 G_{\mu\nu}=G_{\mu\nu}^{(0)}\equiv\delta_{\mu\nu}, \,\,
 \Phi=\Phi^{(0)}\equiv {1 \over 2}Q\phi, \,\, T=0, \label{eq:t4}
\eeq
where $Q=\sqrt{\tilde{k}}$. Then,
$S_{\rm 2d}$ can be written as
\begin{equation}
      S_{\rm 2d}={1 \over 8\pi}\int\,d^2z\sqrt{\hg}
            \left[\hg^{\alpha\beta}
              \partial_\alpha X^{\mu}\partial_\beta X_{\mu}
           +Q{\hat R}\phi \right]
           +\hat{S}_{\rm gh}, \label{eq:t5}
\end{equation}
Around the vacuum \rf{eq:t4}, we make the mode expansion
of the fields $X^{\mu}$. 
Differently from the critical string case, we should be careful about
the existence of the
Liouville mode $\phi=X^0$ which couples to the background charge.
For the mode expansion of this field, the Coulomb gas picture is 
applied \cite{dot} and we obtain,
\bea
 X^{\mu}(z,\bar{z})&=&\varphi^{\mu}(z)+{\bar \varphi}^{\mu}(\bar{z}),
                      \label{eq:28} \\
        \partial \varphi^{\mu}(z)&=&-i\sum_m \alpha_m^{\mu}z^{-m-1},
                      \label{eq:29} \\
   \bar{\partial}{\bar \varphi}^{\mu}(\bar{z})
                   &=&-i\sum_m \bar{\alpha}_m^{\mu}\bar{z}^{-m-1},
                      \label{eq:30}
\eea
where $z=\exp (\tau+i\sigma)$ and
\beq
  [\alpha_m^{\mu}, \alpha_n^{\nu}]=\delta^{\mu\nu}m\delta_{m+n,0}.
                     \label{eq:31}
\eeq
The vacuum for these bosonic fields is defined as,
\beq
 \alpha_m^{\mu}|0>=\bar{\alpha}_m^{\mu}|0>=\left\{\begin{array}{ll}
             0    & \mbox{for $m>0$ and $m=0$, $\mu=i>0$} \\
             -{i \over 2}Q & \mbox{for $m=\mu=0$}\end{array}\right .
                     \label{eq:34}
\eeq

The ghost fields are expanded as
\bea
        c(z)&=&\sum_m c_m z^{-m+1},
                      \nonumber \\
        b(z)&=&\sum_m b_m z^{-m-2}, \, \qquad\qquad
        \{c_n,b_m\}=\delta_{n+m,0}. \label{eq:38}
\eea
Similar formula are obtained for $\bar{b}(\bar{z})$ and
$\bar{c}(\bar{z})$. The vacuum of the ghost would be given in the
next section. And the stress tensors for each field
are obtained as follows,
\bea
     T^{\phi}(z)&=&-{1 \over 2}:\partial\phi\partial\phi:
                       +{Q \over 2}\partial^2\phi,  \label{eq:40} \\
     T^X(z)&=&\sum_{i=1}^c -{1 \over 2}:\partial X^i \partial X^i :,
                       \label{eq:41} \\
     T^{bc}(z)&=& :c\partial b+2\partial c b:.
                       \label{eq:42}
\eea
These formula are used to construct the boundary state in the next section.

\section*{3. Boundary State}\scs{3}

The higher genus manifold is constructed by connecting the one of lower genus
in terms of tubes of the closed string. Then the tube with
the boundary on the surface can be regarded as the elementaly stuff of a
complicated 2d manifold. We construct this tube state
in terms of the loop of the open string whose end point is sewing 
the boundary of the world surface. Another end point disappears in the vacuum
or couples to the external string configurations.
The boundary state with a tube made in this way
is not confromal invariant and it leads to a conformal anomaly.
We considr this formulation for
the noncritical string case given above.

First, consider the boundary, which is set at time $\tau$, where the
following conditions for the end point of the open string 
are demanded,
\bea
    \left.{\partial X_i \over \partial\tau}\right |_{\tau}&=&0, \qquad
            \rm{for} \qquad \rm{i} =1 \sim c,   \label{eq:21} \\
 \left.{\partial \phi\over \partial\tau}\right|_{\tau}&=&-{i \over 2}Q,
                              \label{eq:22}
\eea
\noindent where $\phi=X^0$ and
\beq
    Q=\sqrt{25-c \over 3}, \label{eq:23}
\eeq
as given above.
The second condition \rf{eq:22} for the Liouville fieldis taken such that
it is consistent with \rf{eq:34}. And this implies a definite
form of the noncritical open string action since these boundary conditions
should be derived from the effective action of the noncritical open string.
The Liouville action for the open string has firstly given
by \cite{Dur}, and a more desirable action being consistent with
the idea of \cite{ddk} can be obtained as follows \cite{mans},
\beq
 S_{\phi}^{\rm eff}=S_{\rm L}
       +{1 \over 4\pi}\int_{\partial M}d\hat{s}K_{\hat{g}}Q\phi , 
                       \label{eq:25}
\eeq
where $K_{\hat{g}}$ denotes the extrinsic curvature on the boundary
$\partial M$ and
\begin{equation}
      S_{\rm L}={1 \over 4\pi}\int\,d^2z\sqrt{\hg}
            \left[ {1 \over 2}\hg^{\alpha\beta}
              \partial_\alpha \phi\partial_\beta \phi
           +{1 \over 2}Q{\hat R}\phi \right]. \label{eq:26}
\end{equation}
We notice here that 
$S_L$ is common to the closed string case and the background charge
on the boundary is twice the one on the surface. The second point is 
needed by the consistency with the Gauss-Bonnet theorem. The condition
\rf{eq:22} is obtained if we choose the fiducial metric such as
$\hg_{\alpha\beta}=\delta_{\alpha\beta}$e$^{\varphi}$ and 
$\partial_n\varphi=-i/2$. The scalar ($X^i$) part has no
boundary term. Then the total effective action for the open string 
is written as,
\begin{equation}
      S^{\rm eff}={1 \over 4\pi}\int\,d^2z\sqrt{\hg}
            \left[ {1 \over 2}\hg^{\alpha\beta}
              \partial_\alpha X^i\partial_\beta X^i\right]
             +S_{\phi}^{\rm eff}. \label{eq:27}
\end{equation}

Then the conditions \rf{eq:21} and \rf{eq:22}
lead to the following operator relations,
\bea
 \alpha_{-m}^{\mu}=-\bar{\alpha}_m^{\mu}{\rm e}^{-2m\tau} \qquad 
        \rm{for} \,\, m\neq 0 \, , \nonumber \\
 \alpha_0^i=\bar{\alpha}_0^i=0, \qquad
 \alpha_0^0=\bar{\alpha}_0^0=-{i \over 2}Q \,. \label{eq:32}
\eea
The boundary state which is consistent with the above equations
is obtained as,
\beq
 |B>_{\rm boson}=\exp (\sum_{m=1}^{\infty}\rm{e}^{2m\tau}\alpha_m^{\mu}
                   \bar{\alpha}_m^{\mu})|0>, \label{eq:33}
\eeq
where the bosonic vacuum is defined in \rf{eq:34}.

The constraints for the ghost operators, $c_m$ and $b_m$, are derived from
the requirement of the BRST invariance of the boundary state $|B>$ 
\cite{clny}. The BRST operator is defined as 
\bea
     d&=&{1 \over 2\pi i}\oint J(z)dz, \nonumber \\
     J(z)&=&:[T^{\phi}(z)+T^X(z)+{1 \over 2}T^{bc}(z)]c(z):,
                \label{eq:39} 
\eea
where the stress tensors are given in \rf{eq:40} $\sim$ \rf{eq:42}. And
$\bar{d}$ can be obtained similarly. The BRST invariance for the boundary
state is represented as
\beq
  (d+\bar{d})|B>=0, \label{eq:43}
\eeq 
and this is satisfied for 
\beq
  d= -\bar{d}. \label{eq:44}
\eeq 
Denoting the explicit form of the operator $d$ as
\bea
     d&=& \sum_n (L_n^{\phi}+L_n^X)c_{-n}
              -{1 \over 2}\sum_{n,m}(n-m):c_{-n}c_{-m}b_{n+m}:-c_0
                                       ,  \label{eq:45} \\
     L_n^{\phi}&=&{1 \over 2}\sum_m :a_m^0a_{n-m}^0:
                      +{i \over 2}Q(n+1)a_n^0,
                       \label{eq:46} \\
     L_n^X&=& {1 \over 2}\sum_{i,m} :a_m^ia_{n-m}^i:\ ,
                       \label{eq:47}
\eea
and noticing the relation $L_{-n}=\bar{L}_n$, we find that
\rf{eq:44} is realized by
\beq
  c_n=-\bar{c}_{-n}, \,\,\qquad b_n=\bar{b}_{-n}. \label{eq:48}
\eeq
Then the boundary state with respect to the ghost can be obtained as
\beq
 |B>_{gh}=\exp\left\{\sum_{n=1}^{\infty}{\rm e}^{2n\tau}
        [\bar{c}_{-n}b_{-n}+c_{-n}\bar{b}_{-n}]\right\}
         (c_0+\bar{c}_0)|\downarrow\downarrow>, \label{eq:49}
\eeq
where $|\downarrow\downarrow>$ is the Siegel vacuum which satisfies
\beq
    \begin{array}{ll}
        b_n|\downarrow\downarrow>=0 & \mbox{for $n\geq 0$} \\
        c_n|\downarrow\downarrow>=0 & \mbox{for $n\geq 1$} \\
        <\uparrow\uparrow|b_n=0     & \mbox{for $n\leq -1$} \\
        <\uparrow\uparrow|c_n=0     & \mbox{for $n\leq 0$} \\
                                        \end{array} ,
                     \label{eq:50}
\eeq
and $<\uparrow\uparrow|\downarrow\downarrow>=1$. 
We can find the relations \rf{eq:48} for $n>0$ from the above right
eigenvector. The relations for $n<0$ are found from the following
left eigen vector,
\beq
  <\uparrow\uparrow|(b_0-\bar{b}_0)
       \exp\left\{\sum_{n=1}^{\infty}{\rm e}^{2n\tau}
        [\bar{c}_{n}b_{n}+c_{n}\bar{b}_{n}]\right\}
                         , \label{eq:51}
\eeq

However, we find the right and left cylinder vacuums are orthogonal,
\beq
  <\uparrow\uparrow|(b_0-\bar{b}_0)
         (c_0+\bar{c}_0)|\downarrow\downarrow>=0. \label{eq:52}
\eeq
Then we must demand that
the propagator should be accompanied with the zero modes of $b$ and/or
$c$ in order to get a nonzero cylinder amplitude which is sandwiched by the
boundary states. The appropriate zero mode which is attached to
the propagator of the closed string is the following combination
\cite{clny}, 
$-(b_0+\bar{b}_0)$.
Finally, we arrive at the following boundary state with a tube of the closed
string, 
\beq
   |\Psi>_{\rm B}=-(b_0+\bar{b}_0)
  [L_0+\bar{L}_0-2]^{-1}\left\{|B>+|C>\right\}, \label{eq:53}
\eeq
where $|C>$ denotes the boundary state of Mobius strip which gives
a different normalization coefficient
from that of the annuls $|B>$. This point
is not important here, but the details is seen in \cite{clny}.

It is easy to see that the state \rf{eq:53} leads to the anomaly.
Using the relation,
\beq
     [d+\bar{d}, b_0+\bar{b}_0]=L_0+\bar{L}_0-2 \ , \label{eq:55}
\eeq
we obtain
\beq
   (d+\bar{d})|\Psi>_B=|B>_0+|C>_0, \label{eq:54}
\eeq
where the index $0$ in $|B>_0+|C>_0$ denotes the zero-mass closed string
state in $|B>$ and $|C>$.
For the case of the super-symmetric critical string, the massless states
are gravitons and the dilaton, so the anomaly \rf{eq:54} provides
the modified equations of motion for these fields. However,
we here concentrate on the case of the non-critical string.
In this case, the tachyon is the ground state and its mass becomes imaginary
for $c>1$, and this implies that
the amplitude diverges badly and the anomaly is
not well defined. Then we consider the case of $c=1$, where tachyon is
massless. Then we obtain
\beq
   |B>_0+|C>_0=\kappa(c_0+\bar{c}_0)|\downarrow\downarrow>, \label{eq:56}
\eeq
where $\kappa$ denotes the product of the
string coupling constant and the numerical
factor depending on the details of the open string model considered here.
But its precise value is not necessary here, so we do not consider on
this point. We assume only $\kappa<<1$, because we are considering in
the small coupling region where perturbation with respect to the
string loop expansion is valid.

\section*{4. Equation of Motion}\scs{4}

  The conditions of conformal invariance of \rf{eq:t1}
are summarized by the target space action \rf{eq:t2} for the case
without the string loop corrections. As shown in the previous
section, string loop corrections provide new kind of conformal
anomaly. These anomalies do not cancel within themselves except
for a special case of the superstring model.
According to \cite{clny}, we assume here the two contributions,
tree and the loop, cancel the anomaly each other. Then the target
space action should be modified by adding the term
coming from string loop corrections for getting
more accurate equations.

 In order to obtain the appropriate
effective action, we derive the equations of string
fields according to the operator formalism for the non-critical
string of $c=1$. Consider two kinds of fluctuations around the 
dilaton vacuum. One is the local field fluctuations on the surface, and
it is denoted by $|\Psi>_{\rm T}$,
\beq
   |\Psi>_{\rm T}=\left\{T(x)+h_{\mu\nu}(x)\bar{\alpha}^{\mu}_{-1}
          \alpha^{\nu}_{-1}+\tilde{\Phi}(x)[\bar{c}_{-1}b_{-1}
           +c_{-1}\bar{b}_{-1}]\right\}|\downarrow\downarrow>. \label{eq:e4}
\eeq
The other is the boudary state with the tube
of the closed string which is obtained in the previous section and is denoted
by $|\Psi>_B$ (see \rf{eq:53}). Then we consider the superposition of
these state,
\beq
   |\Psi>=|\Psi>_{\rm T}+|\Psi>_{\rm B}\ . \label{eq:e3}
\eeq
And, the following equations are obtained by imposing the BRST invariance
on this state,
\bea
     (d+\bar{d})|\Psi>&=& \Bigg[
          ({1 \over 2}p^2+{1 \over 8}Q^2-1)T
            +({1 \over 2}p^2+{1 \over 8}Q^2)h_{\mu\nu}
                    \bar{\alpha}^{\mu}_{-1}\alpha^{\nu}_{-1} \nonumber \\
     &+&({1 \over 2}p^2+{1 \over 8}Q^2)
   \tilde{\Phi} (\bar{c}_{-1}b_{-1}+c_{-1}\bar{b}_{-1})\Bigg]
                (c_0+\bar{c}_0)|\downarrow\downarrow> \nonumber \\
      &+&\Bigg[
        [p^{\mu}h_{\mu\nu}+{i \over 2}Qh_{0\nu}-p_{\nu}\tilde{\Phi}]
    \alpha^{\nu}_{-1}\bar{c}_{-1}
                +{i \over 2}Q\tilde{\Phi}\alpha^0_{-1}\bar{c}_{-1} \nonumber \\
 &+&[p^{\nu}h_{\mu\nu}+{i \over 2}Qh_{\mu 0}-p_{\mu}\tilde{\Phi}]
    \bar{\alpha}^{\nu}_{-1}c_{-1}
                +{i \over 2}Q\tilde{\Phi}\bar{\alpha}^0_{-1}c_{-1}
  \Bigg]|\downarrow\downarrow> \nonumber \\
 &+& \kappa(c_0+\bar{c}_0)|\downarrow\downarrow> \, \  =0, 
                       \label{eq:e5}
\eea
From this formula, we obtain the following equations,
\bea
   (-\bar{\nabla}^2+{1 \over 4}Q^2-2)T&=&\kappa, \label{eq:e6} \\
   (-\bar{\nabla}^2+{1 \over 4}Q^2)h_{\mu\nu}&=&
         (-\bar{\nabla}^2+{1 \over 4}Q^2)\tilde{\Phi}=0. \label{eq:e7}
\eea
where $\bar{\nabla}^2=\sum_{i=1}^c\partial_i^2+(\partial_0-Q/2)^2$ and
the gauge fixing conditions,
\beq
    \partial^{\mu}h_{\mu\nu}=Qh_{0\nu}+\partial_{\nu}\tilde{\Phi} .
                       \label{eq:e8}
\eeq
Here we should notice that the zeroth component of the momentum in
\rf{eq:e5} has the following correspondence, $p^0=\alpha^0+iQ/2$.
On the other hand, the differential operator has the correspondence, 
$\partial^0=i\alpha^0$. 

These equations are obtained from the following target space
action, 
\beq
 S^{\rm eff}_{\rm T}=S_{\rm T}
        +2\kappa {1 \over 4\pi}\int d^dX\sqrt{G}{\rm e}^{-\Phi}
           \tilde{T}, \label{eq:e9}
\eeq
where $S_{\rm T}$ is given in \rf{eq:t2} and
\beq
    \tilde{T}=\rm{exp}(\Phi^{(0)})T. \label{eq:e10}
\eeq
Here $\Phi^{(0)}$ is given in \rf{eq:t4}. 
It should be noticed that the added part in $S^{\rm eff}_{T}$ is dependent
on the tree vacuum ($\Phi^{(0)}$). This is reasonable since the boundary
state and the string propagator are intimately related to the background
configurations.

In fact, the equations \rf{eq:e6} and
\rf{eq:e7} can be obtained by the following equations, which are derived
from $S^{\rm eff}_{\rm T}$,
\bea
\nabla^2T-2\nabla\Phi \nabla T&=&{1 \over 2}v'(T)
       +\kappa \rm{e}^{\Phi-\Phi^{(0)}},\label{eq:e12} \\
 \nabla^2\Phi-2(\nabla\Phi)^2&=&-{\tilde{k} \over 2}
        +{1 \over 2}v(T)-{2+d \over 8}\kappa\rm{e}^{\Phi}\tilde{T}, 
                          \label{eq:e13} \\
 -R_{\mu\nu}= -2\nabla_{\mu}\nabla_{\nu} \Phi &+& \nabla_{\mu}T\nabla_{\nu}T
              +{\kappa \over 4}\rm{e}^{\Phi}\tilde{T}G_{\mu\nu}.
                                     \label{eq:e14}
\eea
\noindent where $v'=dv/dT$, and
$\nabla_{\m}$ denotes the covariant derivative with 
respect to the metric $G_{ab}$. We can solve these equations by
keeping the terms of order $\kappa$ and assuming $T$ is small.
The result is given by
\beq
 G_{\mu\nu}=\delta_{\mu\nu}, \,\,
 \Phi={1 \over 2}Q\phi, \,\, T=T^{(1)}\equiv {1 \over 2}\kappa \ . \label{eq:e15}
\eeq
The value of $T$ is shifted by the order of $\kappa$. This solution
implies the occurence of the
condensation of the tachyon due to the string loop corrections.
On the other hand, $G_{\mu\nu}$ and $\Phi$ do not
get any $O(\kappa)$ corrections. This is 
understood from eqs.\rf{eq:e13},\rf{eq:e14}. 

\section*{5. Stability of 2d Surface}\scs{5}

As is well-known, the vacuum obtained for $\kappa=0$
has been unstable for $c>1$. This is because of the appearence of the true
tachyon with negative mass-square. In \cite{coop}, a proposal has been given
to resolve this difficulty. If we seriously consider the term $T^3$ in
the potential $v(T)$, then a non-trivial minimum is found at 
$T(=T_0)=8$ for 
\begin{equation}
     v(T)=-2T^2+T^3/6. \label{eq:t33}
\end{equation}
Then there is no unstable fluctuation at this new vacuum
point, and the authors in \cite{coop} has used this potential to explain
the rolling of the universe from a false vacuum ($T=0$) to
the true one ($T=T_0$).

Althogh this story of the cosmology might be true,
it would be unreasonable to believe the point ($T=T_0$) as
the true vacuum. Because the value $T_0$ is too large to be able to approximate
the potential of $T$ by \rf{eq:t33} near this point.
Around this point, we should need more higher order terms of $T$. Then, it
would be dangerous to say some definite statement in terms of \rf{eq:t33}.
A similar situation is seen in the case of 
the $\lambda\phi^4$ field theory, where
we find a non-trivial minimum of the one-loop corrected
effective potential. However, this minimum could not be related to
the true vacuum since this point
is far from the perturbation. While, a new situation has been opened when the
system is changed to a gauge theory with a complex scalar. 
In this case, the loop
corrections of the gauge fields provides a minimum
within the perturbation. Then this minimum implies
a true vacuum of the theory, and this is known as the
Coleman-Weinberg mechanism \cite{cole}. We can see here
a similar situation for $T$-condensation.

Since $T$ is assumed to be small in obtaining $v(T)$,
the term $T^3$ was neglected in solving \rf{eq:e12} and
we obtained the solution \rf{eq:e15} by including the correction
of $O(\kappa)$. This solution does not have nothing to do with
the minimum of the potential \rf{eq:t33}.
But this does not mean that the term
$T^3$ is meaningless, because this term plays an important role
when we study the stability of the vacuum which is corrected by
$O(\kappa)$.

A way to see the stability of the surface
is to examine the dressed factor of the 
perturbation for the vacuum as done in the case of $\kappa$ = 0, where
the perturbation of the tachyon has been studied. 
Here the following perturbations are considered,
\beq
 G_{\mu\nu}=\delta_{\mu\nu}+h_{\mu\nu}, \,\,
 \Phi=\Phi^{(0)}+\Phi^{(1)}, \,\, T=
               {1 \over 2}\kappa+\tilde{t}. \label{eq:s1}
\eeq
Then, the next equations linearized
with respect to $h_{\mu\nu}$, $\phi$ and $\tilde{t}$ are obtained from
eqs.\rf{eq:e12} $\sim$ \rf{eq:e14},
\bea
   (-\bar{\nabla}^2+{1 \over 4}Q^2
                  -[2-{\kappa \over 4}])\tilde{t} &=&
                  \kappa\phi,  \label{eq:s2} \\
   (-\bar{\nabla}^2+{1 \over 4}Q^2)h_{\mu\nu} &=&
                \kappa\delta_{\mu\nu}\tilde{t},  \label{eq:s3} \\
   (-\bar{\nabla}^2+{1 \over 4}Q^2)\Phi^{(1)} &=&
                 {10+d \over 4}\kappa\tilde{t}, \label{eq:s4}
\eea
where we have used the same gauge condition with \rf{eq:e8}.
In obtaining these equations, we have kept the term up to $O(\kappa)$ 
and neglected the higher order terms like $0(\kappa^2)$.
The above equations, \rf{eq:s2} $\sim$ \rf{eq:s4}, are
different from the case of $\kappa=0$ in the following two points;
(i) The "mass" part of tachyon is shifted from $2$ to
$2-{\kappa \over 4}$ in \rf{eq:s2}, and this comes from the expansion
of $T^3$ in $v(T)$. Then $T^3$ term is essential to obtaining the "mass
shift" of the tachyon.
(ii) Secondly, the couplings among the 
perturbations of the order $\kappa$ appear. 

Due to the second fact, we must solve these equations simultaneoiusly
in order to obtain the dressed factor of the perturbation $\tilde{t}$.
We can solve these in terms of the following ansats,
\beq
 h_{\mu\nu}=\mu_{h}\delta_{\mu\nu}\rm{e}^{\alpha\phi}, \,\,
 \Phi^{(1)}=\mu_{\Phi}\rm{e}^{\alpha\phi}, \,\,
 \tilde{t}=\mu_{t}\rm{e}^{\alpha\phi}. \label{eq:s11}
\eeq
Then we obtain the following relations,
\beq
 \mu_h = {\kappa \over 2-\kappa/4}\mu_t, \,\,
 \mu_{\Phi}={10+d \over 4}{\kappa \over 2-\kappa/4}\mu_{t}. \label{eq:s5}
\eeq
Then we can see that the right hand side of \rf{eq:s2} is the order of
$\kappa^2$, so we can neglect it in solving \rf{eq:s2} and we obtain

\bea
    \alpha &=& {1 \over 2}(Q-\sqrt{Q^2-8+\kappa}) \nonumber \\
           &=& {1 \over 2}(\sqrt{{25-c \over 3}}
                   -\sqrt{{1-c \over 3}+\kappa}). \label{eq:ss5}
\eea
This is the main result of our calculation. For $\kappa=0$, $\alpha$
becomes complex for $c>1$, but it remains real for $\kappa > 0$
even if $c>1$. However the reality of $\alpha$ is restricted to the
small region of $c$, $1+3\kappa >c$, since $\kappa$ is small. Then
it would be necessary to consider the strong coupling region by some
nonperturbative approach in order to see the possibility of 
extending the region of $c$ where $\alpha$ remains real.

The stability of the system can also be seen by seeing the quadratic
terms of the fluctuations $h_{\mu\nu}$, $\Phi^{(1)}$ and $\tilde{t}$
in \rf{eq:e9}. We can assure that there is no tachyonic mode in it
for $c<1+3\kappa$.

It can be seen that
the above conclusion is sensitive to the term $T^3$ of $v(T)$. So
we should comment on the criticism given before on this term.
One is the possibility of rewriting this term
by the derivative term like $(\partial^2T)^3$ in terms of
the field redefinition ~\cite{tsey,bank} which is related to
the ambiguity of the $\beta$-functions
due to its renormalization scheme dependence. 
Generally speaking, the physical consequences should not be
influenced by the renormalization scheme. But our result
\rf{eq:ss5} could not be obtained if we rewrote the potential.
However, the above potential is not obtained by a systematic
perturbative expansion, $\alpha'$ expansion, and $T^3$ term has been
added as a non-perturbative term. Then
we should be careful in rewriting this potential 
in terms of the field redefinition due to the 
renormalization scheme dependence.
Then we had kept the form of \rf{eq:t3} here.
\par
Another criticism \cite{tsey} is the claim that \rf{eq:t3}
allows the solution, $T=$ const., and
this solution of $T\neq 0$ is in contradiction to the
conformal invariance of \rf{eq:t1}. 
However, the potential $v(T)$ is valid for small $T$, so we could not say
from \rf{eq:t3} the existence of the solution of $T\neq 0$ 
which is consistent with the validity of this potential.
But this potential leads to a nontrivial solution for $T$ if we add the 
string-loop corrections to this.
In this case, $T$ could take a small vacuum value which is consistent
with the assumption of small $T$ expansion. And the problem of the conformal
invariance of the 2d action should be reconsidered by taking into account
not only of the short distance fluctuations of the field on the world sheet
but also of the string loop corrections, so the paradox 
would be resolved.

\section*{6. Concluding remarks}\scs{6}

 Here we have examined 
a special quantum surface fluctuation of two-dimensional space-time
in terms of the loop-correction of a open string in order to see
the vacuum structure of 2d gravity and noncritical string theory.
The technique
to calculate this correction had been developed in the 
critical super-string theory
by constructing the boundary state accompaning a tube of the closed
string.
It is applied here to the non-critical string theory and 2d gravity, and 
this correction leads to the conformal anomaly for
the massless channel of the
closed string states. Then the field equation of the tachyon
is modified for $c=1$ case where tachyon is massless.
As a result, this loop-correction leads to the condensation of
the tachyon. Furthermore, the dressed factor of the perturbations 
around this shifted vacuum becomes real even if $c$ exceeds one. 
However this range of $c$ where the dressed factor remains real is restricted
to a very small region above 1, $c<1+3\kappa$. Since $\kappa$ denotes 
the string coupling constant, it would be necessary to consider some
non-perturbative approach to see whether we can extend the
theory to the region where $c$ is enough large compared to one
and to obtain the noncritical string theory of $c>1$.

\newpage


\begin{thebibliography}{99}
\bibitem{ddk}F. David, Mod. Phys. Lett. A3(1988)1651;
    J. Distler and H. Kawai, Nucl. Phys. B321(1989) 509;
    V. Knizhnik, A. Polyakov and A. Zamolodchikov, Mod. Phys. Lett. A3
    (1988)819.
\bibitem{poly} I.R. Klebanov, I.I.Kogan, and A. Polyakov, Phys. Rev. Lett.
             71(1993) 3243;
           C.Schmidhuber, Nucl. Phys. B404(1993)342;
          J.Ambjorn and K.Ghoroku, Int. J. Mod. Phys. A9(1994)5689;
          K.Ghoroku, Phys. Lett. B357(1995)12.
\bibitem{clty} E.S.Fradkin and A.A.Tseytlin, Nucl. Phys. B261(1985)1;
     C.G.Callan, D.Friedan, E.J.Martinec and M.J.Perry,
    Nucl. Phys. B262(1985)593.
\bibitem{ito} C.Itoi and Y.Watabiki, Phys. Lett. B198(1987)486.
\bibitem{coop}A.Cooper, L.Susskind and L.Thorlacius, Nucl. Phys. 
     B363(1991)132.
\bibitem{tsey} A.A.Tseytlin, Phys. Lett. B264(1991)311.
\bibitem{bank} T.Banks, Nucl. Phys. B361(1991)166.
\bibitem{fis} W.Fishler and L.Susskind, Phys. Lett. B173(1986)262.
\bibitem{fta} E.S.Fradkin and A.A.Tseytlin, Phys. Lett. B163(1985)123;
         A.Abouelsaoud, C.G.Callan, C.R.Nappi and S.A.Yost,
    Nucl. Phys. B280(1987)599;
         C.G.Callan, C.Lovelace, C.R.Nappi and S.A.Yost,
    Nucl. Phys. B288(1987)525.
\bibitem{clny} C.G.Callan, C.Lovelace, C.R.Nappi and S.A.Yost,
    Nucl. Phys. B293(1987)83; ibid, B308(1988)221.
\bibitem{schm} C.Schmidhuber, "D branes actions", PUPT-1995,
     hep-th/9601003.
\bibitem{Dur} B.Durhuus, P.Olessen and J.L.Petersen, 
         Nucl. Phys. B198(1982)157;ibid, B201(1892)176;
        B.Durhuus, H.B.Nielsen, P.Olessen and J.L.Petersen, 
         Nucl. Phys. B196(1982)498.
\bibitem{mans} P.Mansfield, Nucl. Phys. B306(1988)630;
       P.Mansfield and R.Neves, hepth/9605097 May 1996.
\bibitem{dot} V.Dotsenko and V.Fateev, Nucl. Phys. B240(1984)312.
\bibitem{cole} S.Coleman and E.Weinberg, Phys. Rev. {\bf D7}(1973)1888.

\end{thebibliography}
\end{document}